\documentstyle[seceq,epsf]{ptptex}
\pubinfo{}  %Editorial Office use
\preprintnumber[3cm]{%<-- [..]: optional width of preprint # column.
KUNS-1325\\PTPTeX ver.0.8\\ August, 1997}
\title{%        %You can use \\ for explicit line-break
Drell-Yan proton-deuteron asymmetry \\
and polarized light-antiquark distributions}
\author{%       %Use \sc for the family name
Masanori {\sc Miyama}
\footnote{Research Fellow of the Japan Society for the Promotion 
of Science. Information on his research is available at 
http://www-hs.phys.saga-u.ac.jp. E-mail: miyama@cc.saga-u.ac.jp.} 
}

\inst{%         %Affiliation, neglected when [addenda] or [errata]
Department of Physics, Saga University, Saga 840-8502, Japan
}

\recdate{%      %Editorial Office will fill in this.
December 31, 1999}

\abst{%         %this abstract is neglected when [addenda] or [errata]
We discuss the Drell-Yan proton-deuteron ({\it p-d}) asymmetry $R_{pd}$,
which is defined by the proton-proton and proton-deuteron
cross-section ratio
$\Delta_{(T)}\sigma^{pd}/2\Delta_{(T)}\sigma^{pp}$,
and its relation to the polarized light-antiquark flavor asymmetry.
Using a formalism of the polarized {\it pd} Drell-Yan process,
we show that the $R_{pd}$ especially in the large-$x_F$ region
is very useful for finding the flavor asymmetry 
in the longitudinally-polarized and transversity distributions. 
Our results are particularly important to study the flavor asymmetry 
in the transversity distributions because they cannot be measured 
by inclusive deep inelastic scattering and $W$-production process.
}

\begin{document}

%%%%%%%%%%%%%%%%%%%%%%%%%%%%%%%%%%%%%%%%%%%%%%%%%%%%%%%%%%%%%%%%%%%%%%%%%%%%%%%

\pagestyle{empty}

\begin{flushleft}
\Large
{SAGA-HE-159-00
\hfill January 7, 2000}  \\
\end{flushleft}

\vspace{2.6cm}
\begin{center}
 
\LARGE{{\bf Drell-Yan proton-deuteron asymmetry}} \\

\vspace{0.2cm}
\LARGE{{\bf   and polarized light-antiquark distributions}} \\

\vspace{1.5cm}
\LARGE
{M. Miyama $^*$} \\
 
\vspace{0.5cm}
\LARGE
{Department of Physics} \\
 
\LARGE
{Saga University} \\
 
\LARGE
{Saga 840-8502, Japan} \\

\vspace{1.5cm}
 
\LARGE
{Talk given at the Circum-Pan-Pacific RIKEN Symposium} \\

\vspace{0.1cm}

{on ``High Energy Spin Physics"} \\

\vspace{0.1cm}

{RIKEN, Wako, Japan, November 3 -- 6, 1999} \\

\vspace{0.05cm}
{(talk on November 4, 1999) }  \\
 
\end{center}
 
\vspace{0.7cm}
\vfill
 
\noindent
{\rule{6.0cm}{0.1mm}} \\
 
\vspace{-0.3cm}
\normalsize
\noindent
{* Research Fellow of the Japan Society for the Promotion of Science. \\
\ \, Information on his research is available 
at http://www-hs.phys.saga-u.ac.jp. \\ 
\ \, E-mail: miyama@cc.saga-u.ac.jp. }
\\

\vspace{+0.1cm}
\hfill
{\large to be published in proceedings}

\vfill\eject
\setcounter{page}{1}
\pagestyle{myheadings}

%%%%%%%%%%%%%%%%%%%%%%%%%%%%%%%%%%%%%%%%%%%%%%%%%%%%%%%%%%%%%%%%%%%%%%%%%%%%%%%
\maketitle

%\tableofcontents

\makeatletter
\if 0\@prtstyle
\def\asp{.3em} \def\bsp{.26em}
\else
\def\asp{.3em} \def\bsp{.3em}
\fi \makeatother

%%%%%%%%%%%%%%%%%%%%%%%%%%%%%%%%%%%%%%%%%%%%%%%%%%%%%%%%%%%%%%%%%%%%%%%%%%%%%%
\section{Introduction}

Nowadays, flavor asymmetry of light-antiquark distributions is
an established fact in unpolarized distributions \cite{skpr}. 
At first, the unpolarized $\bar u/\bar d$ asymmetry was suggested 
by the NMC finding of the Gottfried-sum-rule violation. 
Then, NA51 and E866 collaborations investigated 
the $\bar u/\bar d$ ratio by measuring Drell-Yan 
proton-deuteron asymmetry. Their results clearly showed
that the $\bar u$ and $\bar d$ distributions are different from 
each other. In particular, E866 data revealed detailed $x$ 
dependence of the $\bar d/\bar u$ ratio. Furthermore, 
semi-inclusive deep inelastic scattering data which were
measured by HERMES also showed the flavor asymmetry.

On the other hand, the flavor asymmetry in polarized distributions 
is totally unknown at this stage although there are some model 
predictions. 
Our research purpose is to study the flavor asymmetry in more detail.
At this stage, the longitudinally-polarized parton distributions 
are mainly investigated by measuring the spin structure function $g_1$. 
However, $g_1$ data are not enough to find the flavor asymmetry. 
It may be possible to get the information about the flavor asymmetry 
from semi-inclusive deep inelastic scattering data \cite{my}
which were measured by SMC and HERMES. However, the precision of 
the present data is not enough to determine whether there exists 
the flavor asymmetry although the results show the tendency 
that the $\Delta\bar u - \Delta\bar d$ becomes positive.
Situation is more serious for another polarized distributions, 
namely transversity distributions, since they cannot be measured 
by inclusive deep inelastic scattering and $W$-production processes 
because of the chiral-odd property.
The $W$ production is expected to provide important information 
about the flavor asymmetry in the unpolarized and 
longitudinally-polarized distributions.
In the light of the present situation, we should study other 
independent processes to get the detailed information 
on the flavor asymmetry and to determine the major mechanism 
for creating the asymmetry. 
In this study, we investigate the method in which we use the
polarized proton-deuteron ({\it pd}) Drell-Yan process with 
{\it pp} Drell-Yan \cite{pdasym}.

In Sec. 2, model studies are explained on the light-antiquark 
flavor asymmetry in longitudinally-polarized and transversity 
distributions $\Delta_{(T)}\bar u - \Delta_{(T)}\bar d$.
Here, $\Delta_{(T)}$ denotes $\Delta$ or $\Delta_T$ for the 
longitudinal-polarized or transversity distribution, respectively.
Then, we discuss the relation between the polarized Drell-Yan 
proton-deuteron ({\it p-d}) asymmetry which is defined by the ratio 
of the polarized {\it pd} Drell-Yan cross section to the proton-proton 
({\it pp}) one $\Delta_{(T)} \sigma_{pd}/2 \Delta_{(T)} \sigma_{pp}$
and the polarized flavor asymmetry in Sec. 3.
Furthermore, we show the numerical results for the flavor asymmetry 
effects on the Drell-Yan {\it p-d} asymmetry in Sec. 4.
Conclusions are given in Sec. 5.

%%%%%%%%%%%%%%%%%%%%%%%%%%%%%%%%%%%%%%%%%%%%%%%%%%%%%%%%%%%%%%%%%%%%%%%%%%%%%
\section{Theoretical predictions on the polarized flavor asymmetry}

In this section, we briefly introduce the present status of 
the model studies on the flavor asymmetry 
$\Delta_{(T)}\bar u - \Delta_{(T)}\bar d$. 
First, as one of the origins of the flavor asymmetry, 
there is a perturbative-QCD contribution. 
Next-to-leading-order (NLO) $Q^2$ evolution 
gives rise to the difference between the $\bar u$ and $\bar d$ 
distributions even if the flavor-symmetric distributions are used 
at initial $Q^2$. However, its effect is not so large as to 
reproduce the measured asymmetry in the unpolarized distributions
if the $Q^2$ evolution is calculated in the perturbative $Q^2$ range.
Therefore, we expect that dominant effects come from 
non-perturbative mechanisms.

We have been studying the flavor asymmetry 
in the polarized distributions 
by using typical models for the unpolarized 
$\bar{u}/\bar{d}$ asymmetry, and we have been also investigating 
its effect on the Drell-Yan spin asymmetry $A_{TT}$ \cite{km}. 
One of the typical models is a meson-cloud model. In this model, 
we calculate the meson-nucleon-baryon (MNB) process in which 
the initial nucleon splits into a virtual meson and a baryon, 
then the virtual photon from lepton interacts with this meson. 
Since the lightest vector meson is $\rho$ meson, 
we investigate a $\rho$-meson contribution to the polarized 
flavor asymmetry. We take into account $\Delta$,
in addition to the proton, as a final state baryon, 
and all the possible $\rho$NB processes are considered. 
Among them, the dominant contribution comes from the process 
with $\rho^+$ meson. 
Because the $\rho^+$ has a valence $\bar d$ quark, this
mechanism contributes to the $\bar d$ excess over $\bar u$.
Note that the $\rho$-meson contributions to the flavor asymmetry
are also studied in Ref. 5.

Another typical model in the unpolarized case is 
Pauli-exclusion-principle model. 
Although this mechanism does not seem to explain 
the whole $\bar u/\bar d$ asymmetry, it is still worth while
discussing the polarized asymmetry. 
This model for the polarized case has already studied in 
Ref. 6. According to the SU(6) quark model,  
each quark-state probabilities
in the spin-up proton are given by 
$u^\uparrow = 5/3$, $u^\downarrow = 1/3$, $d^\uparrow = 1/3$, 
and $d^\downarrow = 2/3$, respectively.
Since the probability of $u^\uparrow$ ($d^\downarrow$)
is much larger than that of $u^\downarrow$ ($d^\uparrow$),
it is more difficult to create $u^\uparrow$ ($d^\downarrow$) 
sea than $u^\downarrow$ ($d^\uparrow$) sea because of 
the Pauli-exclusion principle.
Then, if we assume that the magnitude of the exclusion effect 
is the same as the one in the unpolarized case,
$(u_s^\downarrow - u_s^\uparrow)/(u_v^\uparrow - u_v^\downarrow) =
 (d_s - u_s)/(u_v - d_v)$ and a similar equation for 
$d_s^\uparrow - d_s^\downarrow$, the magnitude of 
$\Delta\bar u$ and $\Delta\bar d$ become
$-$0.13 and $+$0.05, respectively. In this way, we find the 
$\Delta\bar u/\Delta\bar d$ flavor asymmetry from this mechanism.

We numerically calculate the flavor-asymmetric distribution
$\Delta_{(T)}\bar u - \Delta_{(T)}\bar d$ by using
the above model results and actual initial distributions.
As a result, we find that both model predictions have similar tendency 
that the $\Delta_{(T)}\bar u - \Delta_{(T)}\bar d$ becomes negative. 
Furthermore, the meson contribution seems to be smaller than 
that of the exclusion model. 
In addition to these mechanisms, there are also some model 
studies on the polarized flavor asymmetry \cite{model}.
To distinguish these mechanisms, we need detailed experimental 
information on the $\Delta_{(T)}\bar u$ and $\Delta_{(T)}\bar d$ 
asymmetry. This is the reason for investigating the
possibility of finding the flavor asymmetry by the polarized 
{\it pd} Drell-Yan data.

%%%%%%%%%%%%%%%%%%%%%%%%%%%%%%%%%%%%%%%%%%%%%%%%%%%%%%%%%%%%%%%%%%%%%%%%%%%%%
\section{Polarized proton-deuteron Drell-Yan process 
and flavor asymmetry}

Recently, a formalism of the polarized {\it pd} Drell-Yan process 
was completed in Refs. 8 and 9.  
They showed that there are additional spin asymmetries compared 
with a spin-1/2 hadron reaction. 
These new spin asymmetries are related to new spin structure 
in a spin-1 hadron and one of major purposes to investigate
the polarized {\it pd} Drell-Yan process is to study this new spin 
structure.

Here, we briefly comment on this topic although it is not
the major purpose in this work. The spin structure of the spin-1/2 
hadron is relatively well investigated by measuring the structure 
function $g_1$. On the other hand, there is no experimental study
for the spin structure of the spin-1 hadron.
In fact, we know that the new spin structure, namely, 
the polarized tensor distributions can be investigated by 
measuring the tensor structure function $b_1$
in the deep inelastic scattering with unpolarized lepton
off the polarized deuteron. The $b_1$ has not be measured 
at this stage, but it is expected to be measured 
at HERMES in future.
In the {\it pd} Drell-Yan, on the other hand, parton-model analysis 
suggests that only three spin asymmetries $A_{LL}$, $A_{TT}$, 
and $A_{UQ_0}$ remain finite and other asymmetries vanish 
in the leading-twist level. In these asymmetries, 
the $A_{UQ_0}$ is a new one in the spin-1 hadron.
In the parton model, this spin asymmetry is expressed by
\begin{equation}
A_{UQ_0} = \frac{ \sum_a e_a^2 \, 
      \left[ \, q_a(x_1) \, \delta \bar q_a^{\, d}(x_2)
             + \bar q_a(x_1) \, \delta q_a^d(x_2) \, \right] }
                { \sum_a e_a^2 \, 
      \left[ \, q_a(x_1) \, \bar q_a^{\, d}(x_2)
             + \bar q_a(x_1) \, q_a^d(x_2) \, \right] }
\ ,
\end{equation}
where, $\delta q_a^d$ and $\delta\bar q_a^d$ represent the 
quark and antiquark tensor distributions in the deuteron.
The subscript $a$ represents quark flavor,
and $e_a$ is the corresponding quark charge. 
Therefore, we can study the tensor distributions also by measuring 
this spin asymmetry. In particular, $A_{UQ_0}$ has an advantage to 
investigate the antiquark tensor distributions in comparison with 
the deep inelastic scattering.
This topic is very interesting but we do not discuss in this paper.
The details are discussed in Refs. 8 and 9, 
so that the interested reader may read these papers.

For investigating the flavor asymmetry, we use the results
for the spin asymmetries $A_{LL}$ and $A_{TT}$.
Because of the existence of the tensor distribution,
it was not clear whether the polarized {\it pd} Drell-Yan cross sections 
are expressed by the same forms as the {\it pp} ones. 
References 8 and 9 revealed this point. 
From their analysis, the difference between 
the longitudinally-polarized {\it pd} Drell-Yan 
cross sections is given by 
\begin{eqnarray}
\Delta \sigma_{pd} & = & \sigma(\uparrow_L , -1_L) 
                       - \sigma(\uparrow_L , +1_L) \nonumber \\
                   & \propto & \sum_a e_a^2 \, 
             \left[ \, \Delta q_a(x_1) \, \Delta \bar q_a^{\, d}(x_2)
           + \Delta \bar q_a(x_1) \, \Delta q_a^d(x_2) \, \right]
\ ,
\label{csl}
\end{eqnarray}
where the $\uparrow_L$, $+1_L$, and $-1_L$ represent 
the longitudinal polarization and $\sigma(pol_p,pol_d)$
represents the cross section with the proton and deuteron 
polarizations, $pol_p$ and $pol_d$.
The $\Delta q_a^{\, d}$ and $\Delta \bar q_a^d$ are the 
longitudinally-polarized quark and antiquark distributions
in the deuteron. The momentum fractions are given by 
$x_1=\sqrt{\tau}e^{+y}$ and $x_2=\sqrt{\tau}e^{-y}$
in the case of small $P_T$. Here, the $\tau$ is defined by 
$\tau=M_{\mu\mu}^2/s$ with dimuon mass $M_{\mu\mu}$ 
and the dimuon rapidity is given by
$y=(1/2)\ln [(E^{\mu\mu}+P_L^{\mu\mu})/(E^{\mu\mu}-P_L^{\mu\mu})]$.
In the same way, the transversely-polarized cross-section difference
is given by
\begin{eqnarray}
\Delta_T \sigma_{pd} & = & \sigma(\phi_p=0,\phi_d=0)
                         - \sigma(\phi_p=0,\phi_d=\pi)
\nonumber \\
 &  \propto & \sum_a e_a^2 \, 
    \left[ \, \Delta_T q_a(x_1) \, \Delta_T \bar q_a^{\, d}(x_2)
          + \Delta_T \bar q_a(x_1) \, \Delta_T q_a^d(x_2) \, \right]
\ ,
\label{cst}
\end{eqnarray}
where the $\phi$ is the azimuthal angle of a polarization vector.
The $\Delta_T q$ and $\Delta_T \bar q$ are quark and antiquark
transversity distributions.

The {\it pp} Drell-Yan cross sections are given simply by replacing 
the distributions in the deuteron in Eqs.~(\ref{csl}) and (\ref{cst}) 
by the ones in the proton.
We use these equations for investigating the flavor asymmetry
in the polarized distributions.
To study the flavor asymmetry, we define the Drell-Yan 
proton-deuteron ({\it p-d}) asymmetry $R_{pd}$ by 
\begin{eqnarray}
R_{pd} & \equiv & \frac{\Delta_{(T)} \sigma_{pd}}
                  {2 \, \Delta_{(T)} \sigma_{pp}} \nonumber \\
       & = &  \frac{ \sum_a e_a^2 \, 
    \left[ \, \Delta_{(T)} q_a(x_1) \, 
              \Delta_{(T)} \bar q_a^{\, d}(x_2)
            + \Delta_{(T)} \bar q_a(x_1) \, 
              \Delta_{(T)} q_a^d(x_2) \, \right] }
              { 2 \, \sum_a e_a^2 \, 
    \left[ \, \Delta_{(T)} q_a(x_1) \, 
              \Delta_{(T)} \bar q_a(x_2)
            + \Delta_{(T)} \bar q_a(x_1) \, 
              \Delta_{(T)} q_a(x_2) \, \right] }
\ .
\label{rpd}
\end{eqnarray}

First, we show the behavior of $R_{pd}$ in the 
large $x_F$ ($= x_1 - x_2$) limit. 
Because sea-quark distributions in the proton become smaller
than other distributions in this limit, the proton sea-quark terms 
in the numerator and denominator of Eq.~(\ref{rpd}) can be ignored.
In our analysis, we neglect the nuclear effects in the deuteron 
and assume the isospin symmetry. Then, the distributions 
in the deuteron can be written in terms of the distributions 
in the proton as
\begin{equation}
\begin{array}{ccc}
\Delta_{(T)} u^d = \Delta_{(T)} u + \Delta_{(T)} d, \ & 
\Delta_{(T)} d^d = \Delta_{(T)} d + \Delta_{(T)} u, \ &
\Delta_{(T)} s^d= 2 \Delta_{(T)} s, \\
\Delta_{(T)} \bar u^{\, d} = \Delta_{(T)} \bar u 
                           + \Delta_{(T)} \bar d, \ &
\Delta_{(T)} \bar d^{\, d} = \Delta_{(T)} \bar d 
                           + \Delta_{(T)} \bar u, \ &
\Delta_{(T)} \bar s^{\, d} = 2 \Delta_{(T)} \bar s. \\
&
\end{array}
\end{equation}
However, for a precise comparison with future experimental data,
the nuclear corrections should be properly included.
Using these relations, $R_{pd}$ becomes
\begin{equation}
R_{pd} (x_F\rightarrow 1) = 1 
- \frac{ [ \, 4 \, \Delta_{(T)} u_v(x_1) \, 
                        -\Delta_{(T)} d_v(x_1) \, ] \, 
   [ \, \Delta_{(T)} \bar u (x_2) - \Delta_{(T)} \bar d (x_2) \, ]}
       { 8 \, \Delta_{(T)} u_v(x_1) \, \Delta_{(T)} \bar u (x_2) \, 
       + 2 \, \Delta_{(T)} d_v(x_1) \, \Delta_{(T)} \bar d (x_2)}
\ , 
\label{rpdxf1}
\end{equation}
where $x_1\rightarrow 1$ and $x_2\rightarrow 0$.
Because of the $\Delta_{(T)} \bar u - \Delta_{(T)} \bar d$ factor, 
the ratio $R_{pd}$ simply becomes one if the distribution 
$\Delta_{(T)} \bar u$ is equal to $\Delta_{(T)} \bar d$.
If we assume that the valence-quark distributions satisfy
$\Delta_{(T)} u_v (x \rightarrow 1) \gg 
 \Delta_{(T)} d_v (x \rightarrow 1)$, Eq.~(\ref{rpdxf1}) can be
written in a more simple form as
\begin{eqnarray}
R_{pd} (x_F\rightarrow 1) & = & 1 - \left [ \,
 \frac{ \Delta_{(T)} \bar u (x_2) - \Delta_{(T)} \bar d (x_2) }
      { 2 \, \Delta_{(T)} \bar u (x_2) } \, \right ]_{x_2\rightarrow 0}
\nonumber \\
  & = & \frac{1}{2} \, \left [ \, 1 
                 + \frac{\Delta_{(T)} \bar d (x_2)}
                        {\Delta_{(T)} \bar u (x_2)} 
                    \, \right ]_{x_2\rightarrow 0}
\ .
\end{eqnarray}
From this equation, it is clear that $R_{pd}$ becomes larger 
(smaller) than one if the $\Delta_{(T)} \bar u$ distribution is 
negative as suggested by recent parametrizations and 
if the $\Delta_{(T)} \bar u$ distribution is larger (smaller) 
than the $\Delta_{(T)} \bar d$.

Next, we discuss the behavior of $R_{pd}$ in another limit, 
namely $x_F\rightarrow -1$. In this limit, the ratio $R_{pd}$ becomes
\begin{equation}
R_{pd} (x_F\rightarrow -1) = 
  \frac{ [ \, 4 \, \Delta_{(T)} \bar u (x_1) \, 
                        +\Delta_{(T)} \bar d (x_1) \, ] \, 
   [ \, \Delta_{(T)} u_v (x_2) + \Delta_{(T)} d_v (x_2) \, ]}
       { 8 \, \Delta_{(T)} \bar u (x_1) \, \Delta_{(T)} u_v (x_2) \, 
        +2 \, \Delta_{(T)} \bar d (x_1) \, \Delta_{(T)} d_v (x_2)}
\ ,
\end{equation}
where $x_1\rightarrow 0$ and $x_2\rightarrow 1$.
If we assume $\Delta_{(T)} u_v (x \rightarrow 1) \gg 
\Delta_{(T)} d_v (x \rightarrow 1)$, the ratio becomes
\begin{equation}
R_{pd} (x_F\rightarrow -1) = 
     \frac{1}{2} \, \left [ \, 1 
                 + \frac{\Delta_{(T)} \bar d (x_1)}
                   {4 \, \Delta_{(T)} \bar u (x_1)} 
                    \, \right ]_{x_1\rightarrow 0}
\ .
\label{rpdxfm1}
\end{equation}
In this equation, we find the extra factor 4 in comparison with 
the equation for $x_F\rightarrow 1$ limit. Therefore, $R_{pd}$
in this limit is not as sensitive to the 
$\Delta_{(T)} \bar u/\Delta_{(T)} \bar d$ asymmetry as the one
in the large-$x_F$ limit although we can investigate 
the flavor asymmetry also in this limit.
From Eq.~(\ref{rpdxfm1}), if the $\Delta_{(T)} \bar u$ distribution 
is equal to the $\Delta_{(T)} \bar d$, $R_{pd}$ becomes $5/8 = 0.625$ 
and if the $\Delta_{(T)} \bar u$ distribution is larger (smaller) than
the $\Delta_{(T)} \bar d$, the ratio becomes larger (smaller) than
this value.

From these analyses, we find that we can investigate the flavor
asymmetry by measuring the polarized {\it pd} Drell-Yan process
and taking the {\it pd} and {\it pp} cross-section ratio.
In particular, $R_{pd}$ in the large-$x_F$ and small-$x_F$
regions are useful to find a signature for the flavor asymmetry.

%%%%%%%%%%%%%%%%%%%%%%%%%%%%%%%%%%%%%%%%%%%%%%%%%%%%%%%%%%%%%%%%%%%%%%%%%%%%%%
\section{Numerical results}

We numerically calculate the Drell-Yan {\it p-d} asymmetry $R_{pd}$
by using a recent parametrization. In this section, 
we show the numerical results and discuss the flavor asymmetry 
effect on the $R_{pd}$ \cite{pdasym}.
As a initial distributions, we use the 1999 version of the LSS 
(Leader-Sidorov-Stamenov) parametrization \cite{lss} for the 
longitudinally-polarized distributions. The LSS99 distributions are 
given at $Q^2$ = 1 GeV$^2$ by assuming SU(3) flavor-symmetric sea.
For the transversity distributions, we simply assume that they are 
the same as the longitudinal distributions at $Q^2$= 1 GeV$^2$ 
by considering the quark-model predictions.
Furthermore, we take center-of-mass energy $\sqrt{s}=50$ GeV 
and dimuon mass $M_{\mu\mu}=5$ GeV.
Although there are some model predictions for the flavor asymmetry
in the polarized distributions as explained in Sec. 2,
we simply take the $\Delta_{(T)} \bar u / \Delta_{(T)} \bar d$ 
ratio as
\begin{equation}
r_{\bar q} \equiv \frac{\Delta_{(T)} \bar u}{\Delta_{(T)} \bar d}
                = 0.7,\ 1.0,\ {\rm or}\ 1.3,
\end{equation}
at $Q^2$=1 GeV$^2$ in the following analysis.
The initial distributions with these $r_{\bar q}$ are evolved to
those at $Q^2=M_{\mu\mu}^2$ by leading-order (LO) evolution 
equations. We use the FORTRAN programs which are provided in
Ref. 11 for calculating the $Q^2$ evolution of
the longitudinally-polarized and transversity distributions.
Then, the {\it pd/pp} Drell-Yan cross-section ratio $R_{pd}$ is
calculated for each $r_{\bar q}$.
The results are shown in Fig. 1. 
%%%%%%%%%% Figure %%%%%%%%%%%%%%%%%%%%%%%%%%%%%%%%%%%%%%%%%%%
\begin{figure}[b]
  \hspace*{1mm}
  \parbox[t]{\halftext}{
    \epsfxsize = 8.0cm \epsfbox{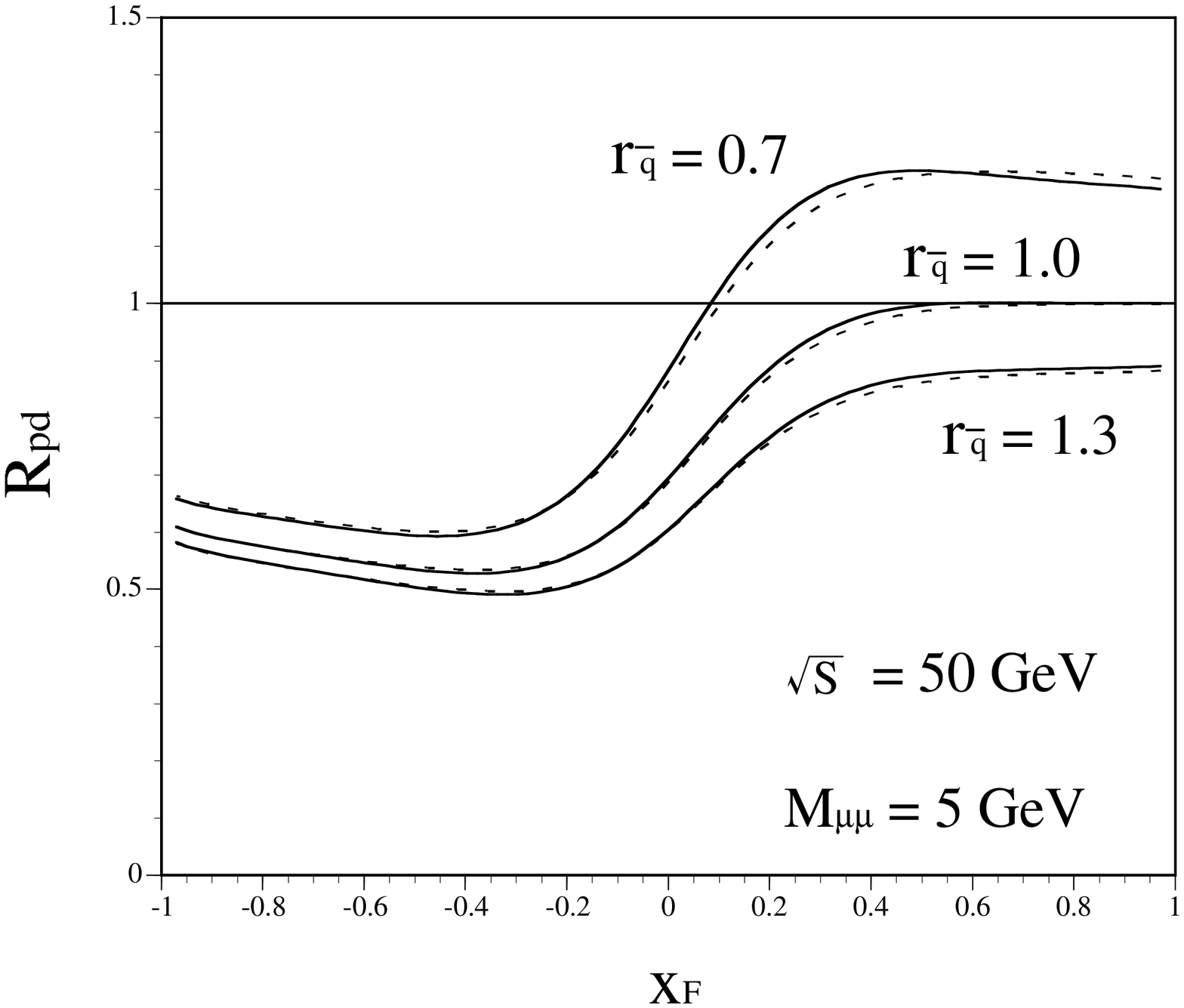}
    \caption{LO evolution results for the $R_{pd}$. 
    The solid (dashed) curves indicate the longitudinally 
    (transversely) polarized ratios (from Ref. 3).}}
  \hspace{-10mm}
  \parbox[t]{\halftext}{
    \epsfxsize = 8.0cm \epsfbox{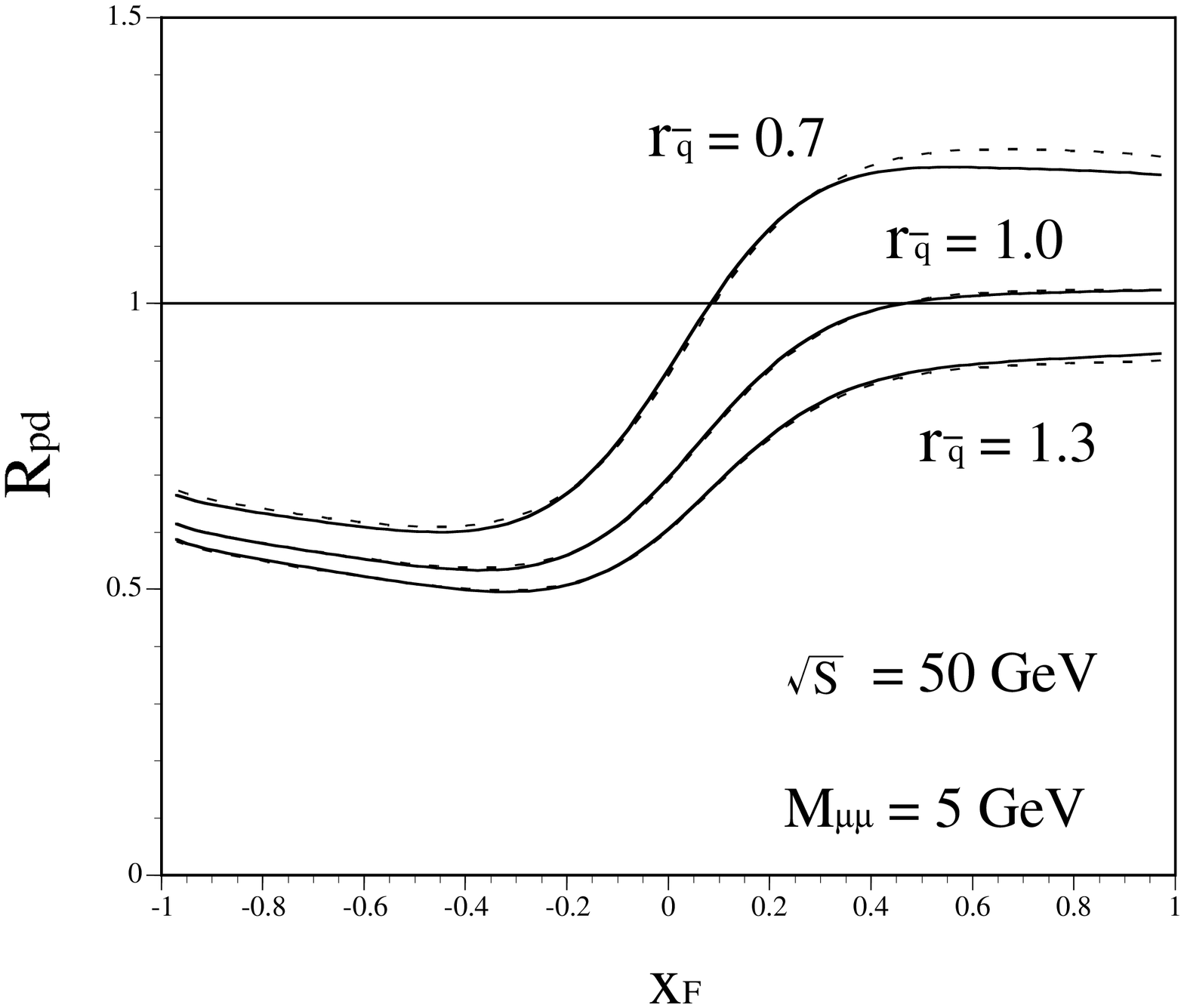}
    \caption{NLO evolution results for the $R_{pd}$. 
    The notations are the same as those in Fig. 1
    (from Ref. 3). }}
\end{figure}
%%%%%%%%%% Figure %%%%%%%%%%%%%%%%%%%%%%%%%%%%%%%%%%%%%%%%%%%
The solid and dashed curves 
represent the longitudinal and transverse results, respectively.
As clearly shown by this figure, the flavor-symmetric 
($r_{\bar q}$ = 1.0) results become one in the large-$x_F$ limit
and 0.625 in the small-$x_F$ limit as discussed in the previous
section. The results with the flavor asymmetry deviate from the
flavor-symmetric ones. In particular, the deviations are 
conspicuous in the large-$x_F$ region. From these results, we find 
that the $R_{pd}$ in the large-$x_F$ region is very useful for
finding the flavor asymmetry in the polarized distributions.
Furthermore, we also find that there is almost no difference
between the longitudinal and transverse results in this
kinematical range if the initial distributions are identical.

In Fig. 2, we show NLO evolution results. 
We evolve the same LSS99 distributions at $Q^2$ = 1 GeV$^2$ 
to those at $Q^2=M_{\mu\mu}^2$ by NLO evolution equations.
The results are almost the same as the LO ones.
However, there is slight deviation. For example, the ratio 
$R_{pd}$ in the large-$x_F$ region is slightly different 
from one although antiquark distributions are flavor symmetric 
at $Q^2$ = 1 GeV$^2$.
It is because the NLO evolution gives rise to the flavor asymmetry.
Although such perturbative-QCD effect is not so large 
in this kinematical region as clearly shown by this figure,
we should include the NLO contributions for a precise analysis.

%%%%%%%%%% Figure %%%%%%%%%%%%%%%%%%%%%%%%%%%%%%%%%%%%%%%%%%%
\begin{figure}[b]
  \hspace*{1mm}
  \parbox[t]{\halftext}{
    \epsfxsize = 8.0cm \epsfbox{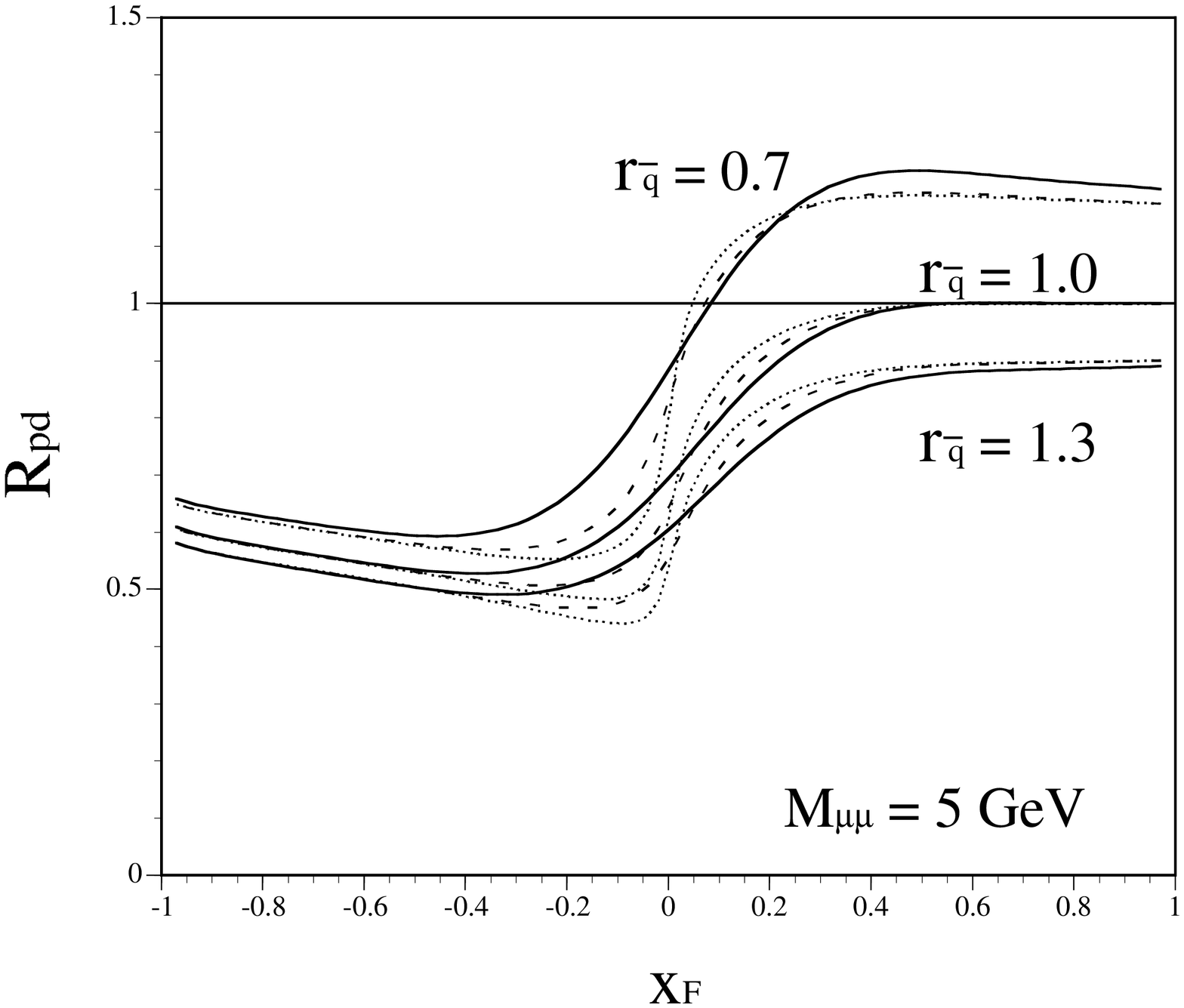}
    \caption{The c.m. energy dependence. The solid, 
    dashed, and dotted curves are the longitudinal ratios 
    at $\sqrt{s}=$50, 200, and 500 GeV, respectively 
    (from Ref. 3).}}
  \hspace{-10mm}
  \parbox[t]{\halftext}{
    \epsfxsize = 8.0cm \epsfbox{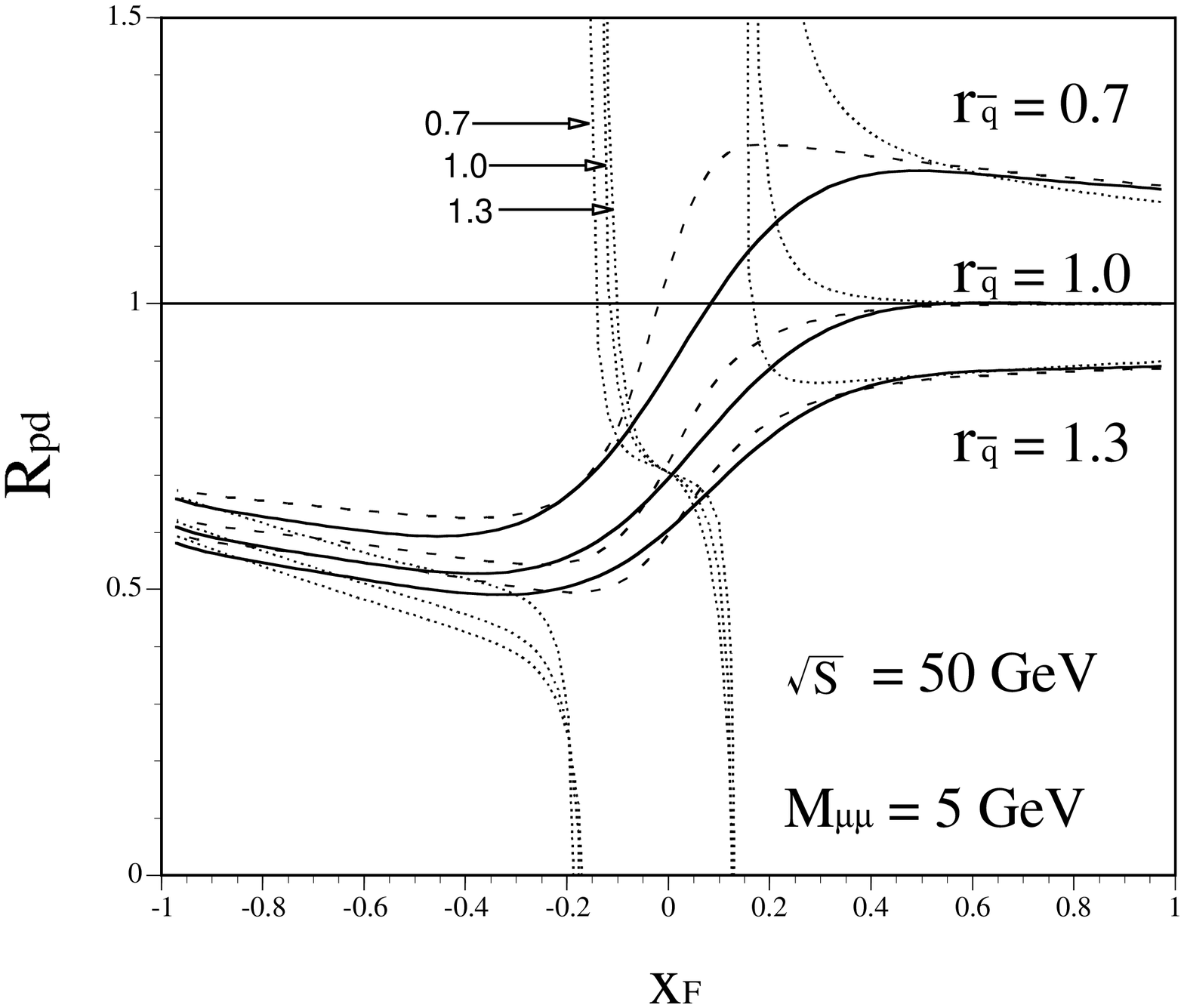}
    \caption{The parametrization dependence. The solid, dashed, 
    and dotted curves are the longitudinal ratios with the LSS99, 
    GRSV96, and GS-A, respectively (from Ref. 3).}}
\end{figure}
%%%%%%%%%% Figure %%%%%%%%%%%%%%%%%%%%%%%%%%%%%%%%%%%%%%%%%%%

Next, we discuss the dependence on the center-of-mass energy $\sqrt s$.
The results for the longitudinal ratio at the RHIC energies 
$\sqrt s$ = 200 and 500 GeV are shown in Fig. 3.
The solid, dashed, and dotted curves indicate the results
at $\sqrt s$ = 50, 200, and 500 GeV, respectively.
The calculated results are almost the same as those 
at $\sqrt s$ = 50 GeV in the large-{} and small-$x_F$ regions. 
However, the ratio in the intermediate-$x_F$ region has much 
dependence on the c.m. energy and becomes a steeper function of $x_F$ 
as $\sqrt s$ increases. From these results, we find that the $R_{pd}$ 
in the intermediate-$x_F$ region is sensitive to the details of
the distributions.

Finally, we show the dependence on the used parametrizations.
In our analysis, we use the LSS99 distributions. In order to 
show the parametrization dependence, we employ the GRSV96 \cite{grsv} 
and the Gehrmann-Stirling set A (GS-A) \cite{gs} distributions. 
The ratios $R_{pd}$ calculated with these initial distributions 
are shown in Fig. 4. The solid, dashed, and dotted curves represent 
the results with LSS99, GRSV96, and GS-A parametrizations.
As shown in this figure, there is not so much difference between
the LSS99 and GRSV results. However, the GS-A results has very
different behavior especially in the intermediate-$x_F$ region.
Because the GS-A antiquark distribution is positive in
the large-$x$ region but becomes negative in the small-$x$ region,
while the LSS99 and GRSV ones are negative in the whole $x$,
the denominator of the $R_{pd}$ becomes zero at certain $x$
and the ratio goes to infinity at such $x$ points. 
Therefore, the $R_{pd}$ in the intermediate-$x_F$ region is
especially useful for determining the detailed $x$ dependence 
of the polarized antiquark distributions.

From these numerical analyses, we find that the $R_{pd}$
is very valuable for investigating the details of the antiquark
distributions.
At this stage, there is no proposal for the polarized {\it pd}
Drell-Yan experiment. However, we think that there are possibilities
at FNAL, HERA, and RHIC.

%%%%%%%%%%%%%%%%%%%%%%%%%%%%%%%%%%%%%%%%%%%%%%%%%%%%%%%%%%%%%%%%%%%%%%%%%%%%%%
\section{Conclusions}

We have studied the Drell-Yan proton-deuteron asymmetry $R_{pd}$
which is defined by {\it pd} and {\it pp} cross-section ratio
$\Delta_{(T)} \sigma_{pd} / 2 \, \Delta_{(T)} \sigma_{pp}$.
Using the formalism for the polarized {\it pd} Drell-Yan process,
we have shown that the $R_{pd}$ is very useful for finding the
light-antiquark flavor asymmetry in the polarized distributions
($\Delta_{(T)} \bar u / \Delta_{(T)} \bar d$), 
especially in the large-$x_F$ region.
We have also shown the dependence on the center-of-mass energy
and the used parametrizations. As a result, we have found
that the $R_{pd}$ in the intermediate-$x_F$ region is
valuable for determining the detailed $x$ dependence of the
polarized antiquark distributions.
Our results are important particularly for the transversity
distributions for which $W$ production does not provide 
information on the $\Delta_T \bar u / \Delta_T \bar d$ asymmetry
because of their chiral-odd nature.

%%%%%%%%%%%%%%%%%%%%%%%%%%%%%%%%%%%%%%%%%%%%%%%%%%%%%%%%%%%%%%%%%%%%%%%%%%%%%%
\section*{Acknowledgements}

This work was partly supported by the Grant-in-Aid for Scientific
Research from the Japanese Ministry of Education, Science, and Culture.
M.M. was supported by a JSPS Research Fellowship for Young Scientists.

%%%%%%%%%%%%%%%%%%%%%%%%%%%%%%%%%%%%%%%%%%%%%%%%%%%%%%%%%%%%%%%%%%%%%%%%%%%%%%

\end{document}